# Assessing the potential of state-of-the-art machine learning and physics-informed machine learning in predicting sea surface temperature

Akshay Sunil[1*], B Deepthi[2], Gaurav Ganjir[3], Muhammed Rashid[3], Rahul Sreedhar[4], Adarsh S[5]

[1]Department of Civil Engineering, Indian Institute of Technology Bombay, Powai, Mumbai 400076, India

[2]Kerala State Council for Science, Technology and Environment, Pattom, Thiruvananthapuram, Kerala 695004

[3]Centre for Climate Studies, Indian Institute of Technology Bombay, Powai, Mumbai 400076, India

[4]Agriculture Insurance Company of India Limited, Ministry of Finance, Government of India, New Delhi, India

[5]Department of Civil Engineering, TKM College of Engineering, Kollam, India

[*]Corresponding author: E-mail: akshaysunil172@gmail.com

## Abstract

The growing adoption of machine learning (ML) in modelling atmospheric and oceanic processes offers a promising alternative to traditional numerical methods. It is essential to benchmark the performance of both ML and physics-informed ML (PINN) models to evaluate their predictive skill, particularly for short- to medium-term forecasting. In this study, we utilize gridded sea surface temperature (SST) data and six atmospheric predictors (cloud cover, relative humidity, solar radiation, surface pressure, u-component of velocity, and v-component of velocity) to capture both spatial and temporal patterns in SST predictions. Four models—Convolutional Neural Network (CNN), Convolutional Neural Network combined with Long Short-Term Memory (ConvLSTM), Transformer, and Transformer with Physics Informed Neural Network (PINN-Transformer)—are assessed for their ability to predict SST at 7-day, 15-day, and 30-day lead times. The accuracy of the models is evaluated using four statistical measures: Anomaly Correlation Coefficient (ACC), Nash-Sutcliffe Efficiency (NSE), Normalized Root Mean Square Error (NRMSE), and Mean Absolute Error (MAE). The results from ACC, NSE, NRMSE, and MAE show that CNN and ConvLSTM models perform well for short-term predictions, with high accuracy in capturing local patterns. At 15-day lead times, the Transformer and the PINN-Transformer models demonstrate superior performance over the other models. Most notably, the PINN-Transformer excels in 30-day predictions, highlighting the importance of integrating physical principles with machine learning to enhance predictive accuracy over longer periods. By capturing both spatial and temporal patterns and incorporating physical constraints, the PINN-Transformer improves long-term SST predictions, emphasizing the potential of hybrid models for more reliable ocean and climate forecasting.

# 1.Introduction

Modelling natural phenomena, particularly oceanic and atmospheric processes, has long captivated researchers, driving advancements in our understanding of complex systems and their behaviour. Running simulations of these phenomena typically involves solving Partial Differential Equations (PDEs), which mathematically represent the underlying physical processes governing oceanic and atmospheric systems (Müller and Scheichl, 2014; Bauer et al., 2015). While PDEs are frequently employed for modelling natural processes, their real-time application is hindered by the intricate and non-linear dynamics inherent in these systems, which pose significant computational challenges (Fatichi et al., 2016; Palmer,2019; Shen et al., 2023). In the field of weather and climate forecasts, numerical weather models are widely used to simulate various climatic variables by approximating PDE's (Lynch 2007; Rodwell and Palmer, 2007; Warner, 2011; Hazeleger et al., 2015; Steppeler and Li, 2022). While these models have the ability to simulate natural processes, they can be demanding in terms of computational resources and may not always provide accurate long-term predictions (Prein et al., 2015; Michalakes, 2020). Furthermore, implementing these models for real-time analysis can pose significant challenges, limiting their practical use. These limitations have prevented significant progress in modelling natural phenomena, with most advancements focused on fine-tuning model parameters (Brotzge et.al 2023).

In recent years, there has been a growing popularity in using machine learning and deep learning models to study and understand natural phenomena (Salman et al., 2015; Scher, 2018; Weyn et al., 2019; Chattopadhyay et al., 2020a; Schultz et al., 2021). This is because of the impressive capabilities demonstrated by these data-driven models in uncovering intricate patterns and connections among different factors that impact various processes, all while requiring less computational resources than traditional methods (Solomatine and Ostfeld, 2008; Solomatine et al., 2009; Grover et al., 2015; Rasp et al., 2020). The advancements in neural networks over the years, from multilayer perceptron to other advanced deep learning algorithms capable of solving complex PDEs, have started addressing the limitations of numerical weather models (Maqsood et al., 2004; Rasp and Lerch, 2018; Chattopadhyay et al., 2020b). As a result, they have become indispensable research tools for climatic science and related fields. This is clear from the significant amount of research being conducted in this area (Abhishek et al., 2012; Zaytar and El Amrani, 2016; Schultz et. al 2021; Hess and Boers 2022; Chen et al., 2023; de Burgh-Day and Leeuwenburg, 2023; Nguyen, 2023).

Sea surface temperature is an important climatic indicator that governs many natural processes, having practical significance in almost all climate related fields (Deser et al., 2010; Hartmann, 2015; Chin et al., 2017; Xiao et al., 2019; Cai et al., 2022). Changes occurring in sea surface temperature is in itself a complex natural process and therefore modelling them using neural networks is crucial in the 21st century (Lins et al., 2013; Zhang et al., 2017). The impacts of anthropogenic climate change have become increasingly evident in recent years, with global temperatures rising at an unprecedented rate. Human activities have caused approximately 1.1°C of warming above pre-industrial levels, primarily due to increased greenhouse gas emissions. This warming trend is not only confined to the atmosphere but also significantly affects the world's oceans. About three-quarters of the Earth's surface is covered by oceans, which are important in shaping the global climate (Webster, 1994; Bollmann et al., 2010). Sea surface temperature (SST) has a significant influence on various environmental issues, such as climate change, marine disasters, and ocean acidification, as well as phenomena like ocean currents and the El Niño-Southern Oscillation (Van Aalst, 2006; Karim and Mimura, 2008; Hobday et al., 2018). Consequently, accurately predicting SST is of great practical importance and can greatly benefit numerous environment-related research activities and applications. However, predicting SST is challenging due to its dynamic changes over time and space (O'Carroll et al., 2019).

Deep learning methods are used for SST prediction due to their strong capability in handling temporal and spatial data (Zhang et al., 2017). For instance, the Long Short-Term Memory (LSTM) leverages the temporal dependencies in SST sequences for prediction, while the Convolutional Neural Network (CNN) model treats SST data as a 2-D map, learning spatial dependencies between different regions to make predictions. Recently, studies have combined CNN with LSTM networks to effectively capture both spatial and temporal dependencies in

SST data and is mainly termed as 'ConvLSTM' (Zhang et al., 2020; Qiao et al., 2021; Hao et al., 2023). Later studies have integrated Convolutional layers, LSTM networks, and Transformer architecture which is termed as ConvLSTM-Transformer (Choudhury et al., 2023). This combination captures spatial, sequential, and global interactions, thereby enhancing prediction accuracy and generalization. By effectively handling long-term dependencies, this model has the potential to revolutionize SST forecasting and improve the understanding of oceanic conditions. While data-driven neural networks are continually improving, there is increasing interest in using physics-informed neural networks (PINNs). PINNs take advantage of neural networks ability to approximate functions by directly incorporating the physical laws described by partial differential equations (Raissi et al., 2019; Cai et al., 2021). Studies have used Physics-Informed ConvLSTM model (ConvLSTM-PINN) for predicting SST (Yuan et al., 2023), which incorporates domain-specific physical knowledge into the learning process, ensuring that the model's predictions are not only data-driven but also consistent with established physical laws.

Hence, different models exist for predicting SST, which is crucial due to the significant increase in SST over recent decades. One alarming consequence of higher SSTs is the increased frequency and intensity of cyclones in the Arabian Sea (Evan and Camargo, 2011; Murakami et al., 2017). Historically, the Arabian Sea has experienced fewer cyclones compared to the Bay of Bengal. However, recent data indicates a sharp rise in cyclonic activity in the Arabian Sea, with Cyclone Tej and Cyclone Biparjoy in 2023 being recent examples. Moreover, studies have shown that the number of cyclones in the Arabian Sea has increased by 52% in the past two decades, while the number of very severe cyclonic storms has increased by 150% (Simpkins, 2021). Given the critical importance of SSTs in influencing weather patterns, climate dynamics, and marine ecosystems, accurate SST prediction is essential. Therefore, this study evaluates a range of machine learning models to identify the most accurate for SST prediction across various lead times. Furthermore, the relevance and potential of physics-informed machine learning (PIML) in improving predictive models are supported by recent research, such as Raissi et al. (2019), who showed how physics-informed neural networks (PINNs) can effectively tackle forward and inverse problems involving nonlinear partial differential equations. Similarly, Karniadakis et al. (2021) provide a comprehensive review of the applications of PINNs across various fields of physics and engineering, underscoring the robustness of incorporating physical laws into machine learning frameworks. Hence, in this study, we test four models, starting with a basic Convolutional Neural Network (CNN). The second model integrates CNN with a Long Short-Term Memory (LSTM) network, known as ConvLSTM. The third model utilizes a standard Transformer architecture for sequential data processing, while the fourth model builds on this by incorporating physics-informed constraints, resulting in the PINN-Transformer. These four models are compared based on their performance in predicting SST and by comparing these models, the study aims to provide insights into the effectiveness of integrating physical knowledge and advanced deep learning techniques for improving SST prediction.

In this study, six atmospheric variables (cloud cover, relative humidity, solar radiation, surface pressure, u-component of velocity, and v-component of velocity) from ERA5 reanalysis data are used to predict SST. These atmospheric variables, collected over the period from 2000 to 2023 at 6-hourly intervals, help establish a relationship with SST. The prediction is performed using four models: CNN, ConvLSTM, Transformer, and PINN-Transformer, to identify the most accurate model. The first 17 years of data are used to train the models, and the remaining 7 years are used for testing. The models are trained to predict SST with lead time 7-day, 15-day, and 30-day. The accuracy of the predictions is evaluated using four statistical measures: Anomaly Correlation Coefficient (ACC), Nash-Sutcliffe Efficiency (NSE), Normalized Root Mean Square Error (NRMSE), and Mean Absolute Error (MAE).

## 2. Study Area and Data Used

### 2.1. Study Area

This study focuses on the Arabian Sea, specifically the region extending from 7.25°N to 18.75°N and 62°E to 72°E (as depicted in Figure 1), to assess the performance of the models in predicting sea surface temperature (SST). The Arabian Sea is selected due to its unique and complex oceanographic and atmospheric characteristics, which make it an ideal area for studying SST and related climate phenomena. The region experiences highly dynamic processes driven by the Indian monsoon system, which influences seasonal wind patterns, precipitation, and ocean currents. These features, combined with the area's susceptibility to climate change, make the Arabian Sea a critical region for SST monitoring and climate modelling. The seasonal reversal of wind patterns during the monsoon period significantly impacts the SST, contributing to variability that plays a key role in influencing weather patterns across the Indian subcontinent. For instance, SST fluctuations in the Arabian Sea are closely linked to the intensity and onset of the monsoon, which, in turn, affects agriculture, water resources, and energy systems in the region. In addition to the local weather systems, the

Arabian Sea also acts as a critical component in the broader climate system, influencing global atmospheric circulation patterns. Understanding SST variability here is essential for predicting climate extremes such as cyclones, droughts, and floods. The vulnerability of this region to climate change exacerbates the need for accurate SST predictions, as rising temperatures and changing weather patterns could disrupt marine ecosystems, fisheries, and coastal livelihoods. Consequently, monitoring SST in the Arabian Sea offers insights into a wide range of ecological and socio-economic issues, extending the importance of this research beyond pure climatology.

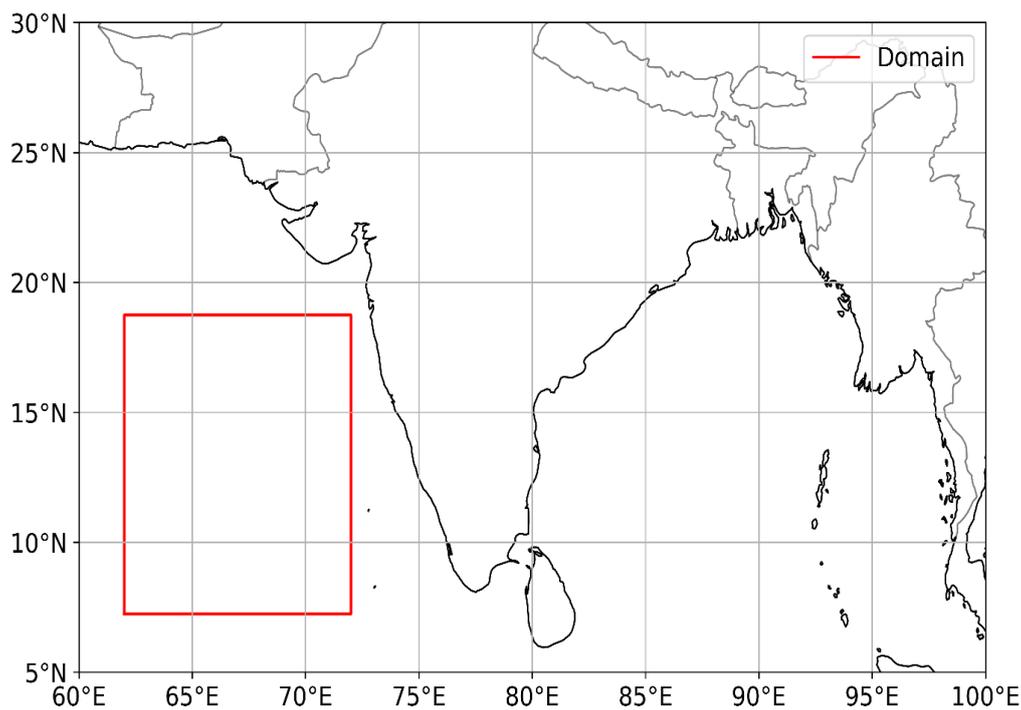

**Figure 1:** The study area in the Arabian Sea, highlighted by the red rectangle, extends from 7.25°N to 18.75°N latitude and 62°E to 72°E longitude. This region, encompassing the western coast of India, parts of Southeast Asia, and the northern Indian Ocean, has been selected for benchmarking the models in this study.

**2.2. Data Used**

In this study, SST predictions are generated using atmospheric variables obtained from the ERA5 reanalysis dataset. ERA5 provides comprehensive climate data, including temperature, wind, and humidity, among others, with a high spatial resolution of 0.25° x 0.25°. This fine-scale resolution is ideal for capturing detailed atmospheric and oceanic interactions; however, it also comes with high computational demands, particularly when running models over large time periods or regions. To optimize computational efficiency while retaining the integrity of the data, bilinear interpolation is applied to adjust the resolution to 0.5° x 0.5°. This method is widely used in climate studies because it effectively smooths and resamples data, making it computationally manageable while preserving the spatial patterns essential for accurate model performance (Vandal et al., 2017; Jose et al., 2022).

In this study, six atmospheric variables—cloud cover, relative humidity, solar radiation, surface pressure, u-component of velocity, and v-component of velocity—are considered to predict SST. Predictor variables are selected by leveraging the understanding of physics, with proper justification for choice of each variable. Bilinear interpolation is applied to adjust the spatial resolution of the selected atmospheric variables and SST to 0.5° x 0.5°. The selected variables are analysed over the period from 2000 to 2023 at 6-hourly time steps to establish their relationship with SST during the same period. Given the 6-hourly time step, there are four values for each day. The dataset spans 24 years, with the first 17 years (2000–2016) used for model training and the remaining 7 years (2017–2023) are used for model testing

Table 1 provides a detailed justification for the chosen atmospheric variables, highlighting their established relationships with sea surface temperature (SST) and supported by relevant literature references. Cloud cover affects solar radiation reaching the ocean surface, where increased cloudiness typically lowers SST due to reduced heating. Relative humidity influences evaporation rates and atmospheric insulation, with higher humidity leading to less evaporative cooling and thus higher SST, while also affecting cloud formation which impacts solar insulation. Solar radiation, as the primary energy source, directly modifies the ocean's heating, with fluctuations significantly altering SST, particularly in high insolation areas. Surface pressure affects SST by altering wind patterns and atmospheric circulation, which in turn influence weather patterns such as clouds and precipitation, affecting both solar radiation and evaporation. Finally, the U-component and V-component of wind play crucial roles in driving ocean currents and heat distribution, where wind patterns can cause variations in SST through processes like upwelling and downwelling, as well as horizontal heat advection.

**Table 1:** Justification for the selection of atmospheric variables

| Variable | Fundamental Principle | Established Relationship | Supporting Literature |
|---|---|---|---|
| Cloud Cover | Affects the amount of solar radiation reaching the ocean surface. | Cloud cover can modulate the incoming solar radiation, affecting the heating and cooling rates of the ocean surface. More cloud cover generally leads to lower SST due to reduced solar heating. | Curry et al. (1993), Wallace and Hobbs (2006) |
| Relative Humidity | Affects evaporation rates and atmospheric insulation. | Higher humidity reduces evaporative cooling, leading to higher SST, while lower humidity enhances cooling. Humidity also influences cloud formation, affecting solar radiation reaching the ocean. | Holton (2004), Curry et al. (1993) |
| Solar Radiation | Primary energy source for Earth's climate system. | Directly influences the heating of the ocean's surface, affecting SST. Variations in solar radiation can cause significant changes in SST, especially in high insolation regions. | Stewart (2008) |
| Surface Pressure | Affects wind patterns and atmospheric circulation. | High and low-pressure systems alter SST by changing weather patterns (e.g., clouds, precipitation) and wind conditions, which affect solar radiation and evaporation. | Houghton (2009), Trenberth and Hurrell (1994) |
| U-Component of Wind | Drives ocean currents and heat distribution. | Wind patterns influence oceanic processes like upwelling and advection, affecting SST. Strong winds can lead to cooler SST through upwelling, while weaker winds can cause warming. | Bakun (1990), Kraus and Businger (1994) |
| V-Component of Wind | Drives ocean currents and heat distribution. | Similar to U-component, wind patterns influence SST through upwelling, downwelling, and horizontal heat advection. | Bakun (1990), Kraus and Businger (1994) |

## 3. Methodology

In this study, we benchmark the performance of four models—CNN, LSTM, Transformer, and PINN-Transformer—for predicting Sea Surface Temperature (SST) with lead times of 7, 15, and 30 days. Each model is used to capture different aspects of SST dynamics: CNN to capture spatial features, LSTM for temporal dependencies, Transformer for long-range attention, and PINN-Transformer to integrate physical laws of ocean dynamics. The models are evaluated based on metrics such as ACC, NSE, NRMSE, and MAE. The performance of the models is compared across the three lead times, providing insights into their suitability for short- and medium-term SST predictions. Each model leverages different computational architectures and underlying mechanisms to predict spatio-temporal changes in SST.

The CNN model is used to capture the spatial features of SST and other climate variables such as cloud cover, relative humidity, solar radiation, surface pressure, u-component of velocity, and v-component of velocity (Ghosh et al., 2020; Tang et al., 2022). The architecture consists of two convolutional layers with LeakyReLU activations, batch normalization for stable training, and dropout layers to prevent overfitting. These convolutional layers extract key spatial patterns from the input data, which are flattened and passed through dense layers to predict SST values. The model is trained using the Mean Squared Error (MSE) loss function and optimized with the Adam optimizer. The learning rate is dynamically adjusted using the Reduce Learning Rate on Plateau (ReduceLROnPlateau) callback to prevent overfitting and enhance model convergence. The ConvLSTM model combines the strengths of convolutional layers and LSTM units to capture both spatial and temporal dependencies in the SST data (Alhussein et al., 2020). It processes sequences of input data over time, allowing the model to learn the temporal dynamics of SST while preserving the spatial structure. The architecture includes two ConvLSTM layers followed by batch normalization and dropout for regularization. The output is passed through dense layers to generate the SST predictions. This model is also trained using MSE loss, with the Adam optimizer, and the learning rate is adjusted using a ReduceLROnPlateau strategy.

The Transformer model is used to capture long-range dependencies in SST data through its self-attention mechanism (Choudhury et al., 2023). Spatial features are first extracted using convolutional layers, which are then passed through multiple Transformer encoder blocks. Each encoder block includes multi-head attention layers and feed-forward layers to model both the spatial and temporal aspects of SST data. The final predictions are made using fully connected layers with linear activation. The model is optimized using the Adam optimizer and trained with MSE loss. Training and validation losses are tracked across epochs to evaluate model performance. The PINN Transformer model integrates physical constraints into the Transformer architecture by incorporating ocean dynamics into the learning process (Raissi et al., 2019). The physics-informed loss function includes terms related to the divergence and Laplacian of the velocity components (u and v), ensuring that the predicted SST values adhere to the laws of physics governing ocean heat dynamics. The model balances traditional prediction loss with the physics-informed loss to improve generalization, especially in regions with limited data. Like the other models, it is trained using the Adam optimizer and MSE loss, with a focus on capturing both spatio-temporal patterns and physical consistency in SST predictions. The steps involved in the methodology are as follows:

**Step 1:** The dataset includes SST and six atmospheric variables: cloud cover, relative humidity, solar radiation, surface pressure, u-component of velocity, and v-component of velocity. These variables are sourced from the ERA5-reanalysis dataset.

**Step 2:** The data are normalized to establish a consistent range across all variables and grids.

**Step 3:** SST time series are modeled at each grid using four different models: CNN, ConvLSTM, Transformer, and PINN-Transformer. For each model, SST is predicted at 7-day, 15-day, and 30-day lead times.

**Step 4:** The accuracy of each model at 7-day, 15-day, and 30-day lead times is evaluated using the ACC, NSE, NRMSE, and MAE metrics.

## 4. Results and Discussion

Sea surface temperature (SST) predictions are made using four models: CNN, ConvLSTM, Transformer, and PINN-Transformer. The performance of each model during the training and testing phases is evaluated using metrics such as ACC, NSE, NRMSE, and MAE. The study assesses the efficiency of these models across various lead times, specifically 7-day, 15-day, and 30-day. The results for each of these lead times are discussed separately below.

### 4.1. Results for 7-day lead time

The spatial resolution of 0.5° × 0.5° results in 504 grid points within the study domain, with SST predictions made at each grid. The ACC, NSE, NRMSE, and MAE values obtained during the testing period are displayed in Figure 4. Each row in the figure represents a different metric, while models are presented across columns. For simplicity, the ConvLSTM model will be referred to as LSTM from now on. The results indicate that the CNN, LSTM, and Transformer models consistently achieve high ACC and NSE values, demonstrating a strong correlation between observed and predicted SST, along with efficient model performance. These models also show low NRMSE and MAE values, indicating accurate and reliable forecasts with minimal errors. Conversely, the PINN-Transformer model exhibits a different trend, characterized by noticeably lower ACC and NSE values, and higher NRMSE and MAE, particularly in certain regions. This suggests that the PINN-Transformer model performs less consistently in predicting SST anomalies at the 7-day lead time, as reflected by its comparatively lower accuracy, efficiency, and higher error rates. The visual representation highlights the varying levels of model effectiveness, emphasizing their relative strengths and weaknesses in short-term SST forecasting.

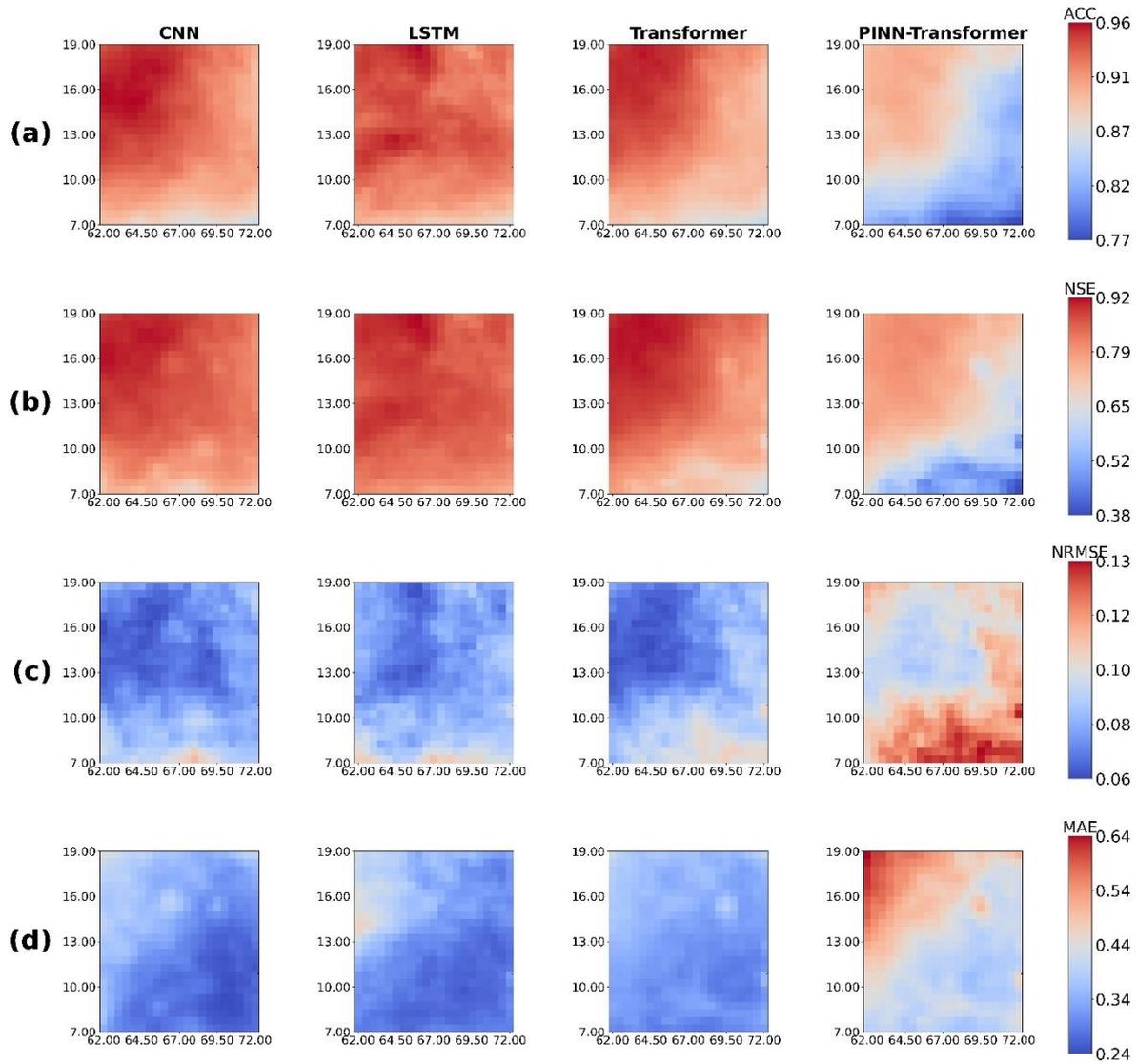

**Figure 4. (a)** CC; **(b)** NSE; **(c)** NRMSE; and **(d)** MAE values obtained during the testing period for CNN, LSTM, Transformer, and PINN-Transformer models for 7-day lead time.

Figure 5 presents scatter plots comparing observed and predicted SST values for four randomly selected grids from the 504 grids analysed in the study. These grids are chosen to illustrate the variation in NSE values across different models. Scatter plots are provided for all four models considered in the study, with the latitude and longitude coordinates of the selected grids indicated at the top of the figure. Each row represents a different model, while the columns correspond to specific grids. The NSE values displayed on each plot measure the accuracy of the models, with higher values indicating better performance. The CNN model (Figure 5(a)) shows NSE values ranging from 0.73 to 0.78, indicating a moderate to good correlation between observed and predicted SST. The LSTM model (Figure 5(b)) exhibits higher NSE values, ranging from 0.77 to 0.82, suggesting greater predictive accuracy and consistency across the grids. The Transformer model (Figure 5(c)) displays NSE values between 0.67 and 0.78, indicating variability in performance but generally good accuracy. In contrast, the PINN-Transformer model (Figure 5(d)) presents the lowest NSE values, ranging from 0.45 to 0.65, indicating less accurate predictions. This model also shows a wider spread of data points around the line of perfect agreement, reflecting greater discrepancies between observed and predicted values. Overall, Figure 5 highlights the varying levels of predictive accuracy among the models, with the CNN and LSTM models providing more reliable predictions, while the PINN-Transformer model demonstrates more variability and lower accuracy.

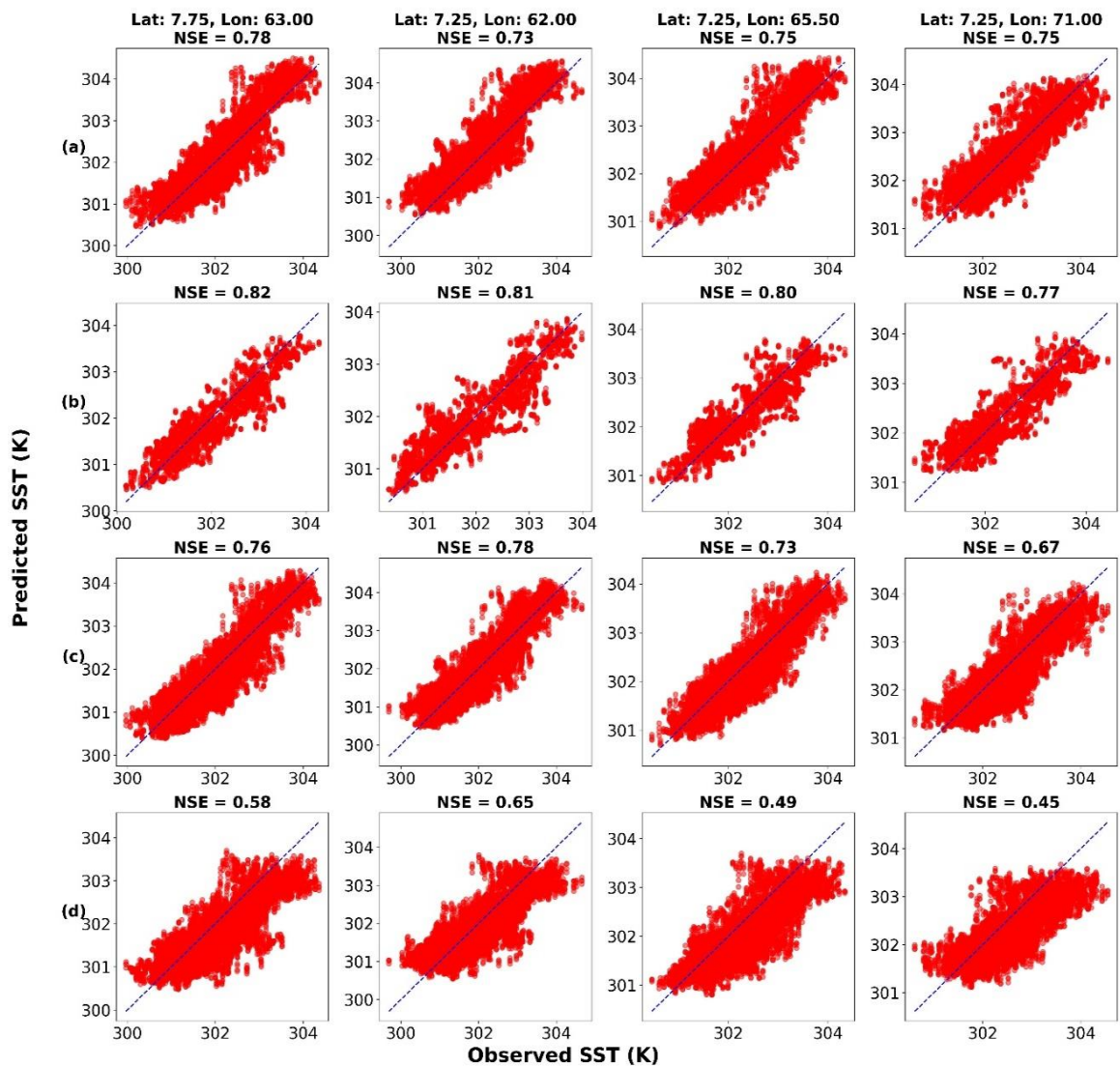

**Figure 5.** Scatter plot between observed and predicted SST for 7-day lead time for different models: **(a)** CNN; **(b)** LSTM; **(c)** Transformer; and **(d)** PINN-Transformer.

## 4.2. Results for 15-day lead time

SST predictions for 15-day lead time are made using four different models, with the ACC, NSE, NRMSE, and MAE values during the testing period are shown in Figure 6. The CNN model (Figure 6(a)) shows high ACC values, especially in the northern regions, indicating strong predictive accuracy. It also maintains relatively high NSE values, reflecting efficient performance, while the NRMSE and MAE metrics indicate generally low error rates, though some variability is observed. The LSTM model (Figure 6(b)) performs similarly to the CNN model, with high ACC and NSE values, indicating good predictive capabilities, albeit with slightly higher NRMSE and MAE values, suggesting marginally lower accuracy. The Transformer model follows a similar pattern, with high ACC and NSE values comparable to those of the CNN and LSTM models; however, the NRMSE and MAE metrics reveal slightly more pronounced errors, indicating some variability in prediction accuracy. Finally, the PINN-Transformer model (Figure 6(d)) shows lower ACC and NSE values, particularly in the southern regions, indicating less accurate predictions. This model also exhibits higher NRMSE and MAE values, reflecting greater errors in its forecasts. Overall, Figure 6 demonstrates that while the CNN, LSTM, and Transformer models generally provide accurate and reliable SST forecasts, the PINN-Transformer model shows more variability and higher error rates, especially in specific regions, suggesting it may be less consistent in its predictive performance.

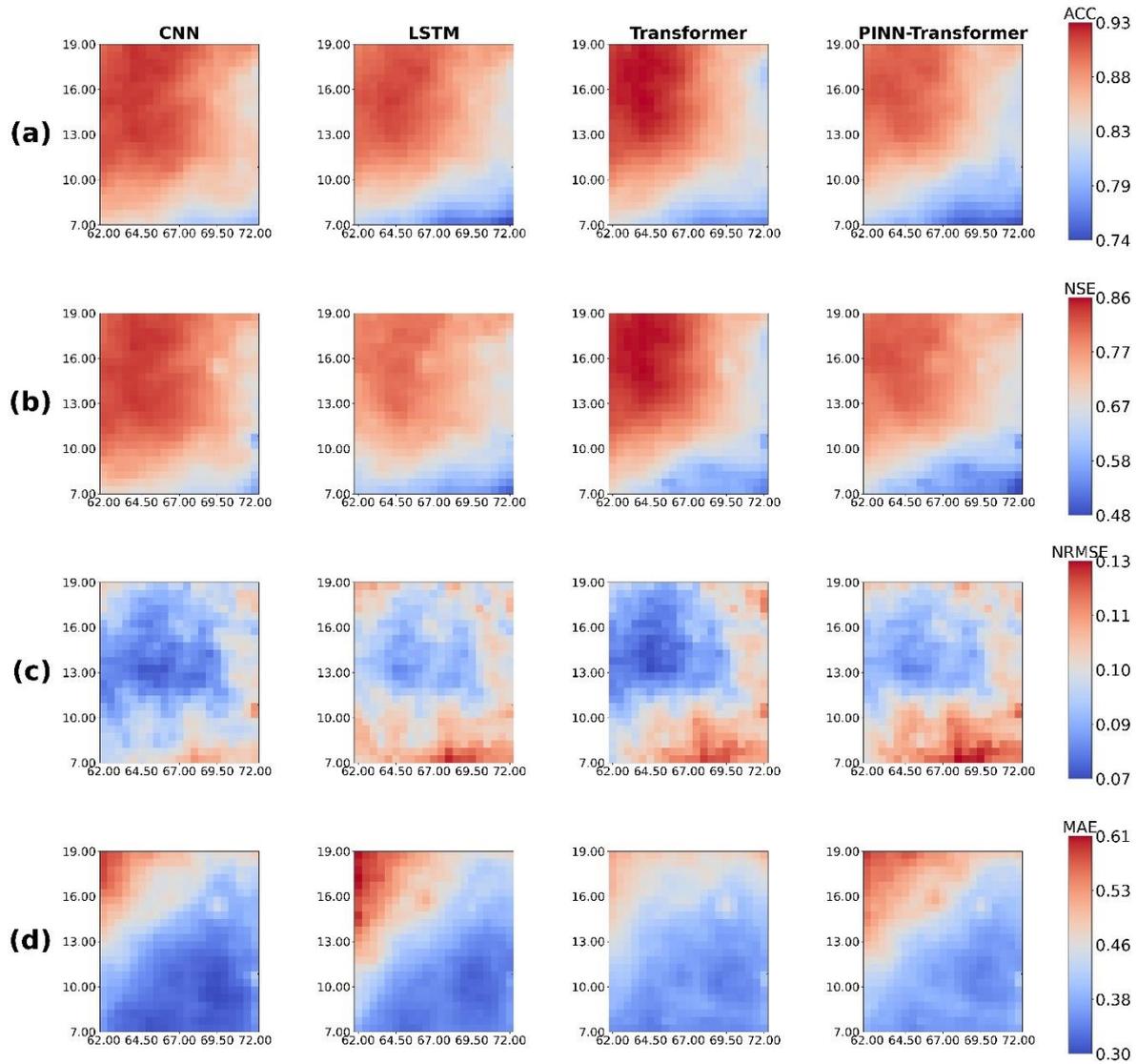

**Figure 6. (a)** CC; **(b)** NSE; **(c)** NRMSE; and **(d)** MAE values obtained during the testing period for CNN, LSTM, Transformer, and PINN-Transformer models for 15-day lead time.

Figure 7 presents scatter plots comparing observed and predicted SST for 15-day lead time across four models: CNN, LSTM, Transformer, and PINN-Transformer. The scatter plots correspond to the same four grids shown in Figure 5, allowing for a comparison of how prediction accuracy varies with increased lead time at these specific locations. Each row represents a different model, while each column corresponds to one of the four grids.

The CNN model (Figure 7(a)) demonstrates moderate predictive accuracy, with NSE values ranging from 0.68 to 0.72, indicating a reasonable correlation between observed and predicted SST. The LSTM model (Figure 7(b)) shows slightly lower NSE values, between 0.54 and 0.64, suggesting moderate predictive accuracy. The Transformer model (Figure 7(c)) displays NSE values from 0.56 to 0.69, reflecting a similar range of performance. The PINN-Transformer model (Figure 7(d)) exhibits the lowest NSE values, ranging from 0.52 to 0.65, indicating less accurate predictions compared to the other models. These scatter plots highlight the variability in model performance, with some models showing better alignment between observed and predicted values, as reflected in their NSE values. Overall, the trend suggests that prediction accuracy varies across different models and geographic grids.

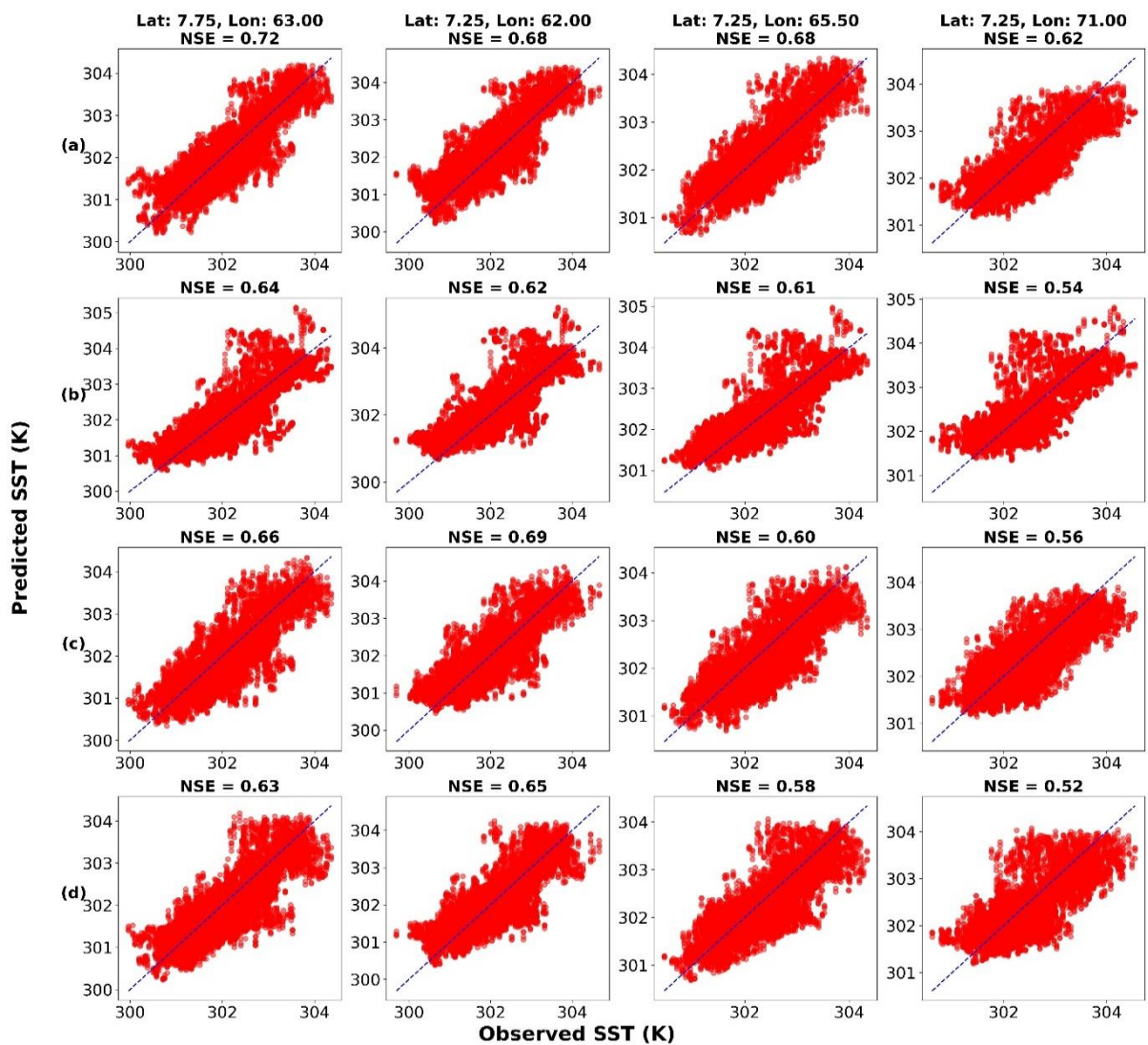

**Figure 7**. Scatter plot between observed and predicted SST for 15-day lead time using different models: **(a)** CNN; **(b)** LSTM; **(c)** Transformer; and **(d)** PINN-Transformer.

### 4.3. Results for 30-day lead time

Figure 8 shows the ACC, NSE, NRMSE, and MAE values obtained during the testing period across four models for 30-day lead time. The CNN and LSTM models display moderate ACC and NSE values, indicating reasonable predictive accuracy, though with some variability in error as reflected by the NRMSE and MAE metrics. The Transformer model performs slightly better, with higher ACC and NSE values in certain regions, though it still shows some variability in error metrics. In contrast, the PINN-Transformer model stands out with the highest ACC and NSE values, demonstrating a strong correlation between observed and predicted SST, along with efficient performance. This model also exhibits the lowest NRMSE and MAE values, indicating fewer errors and more accurate predictions.

The superior performance of the PINN-Transformer model, particularly at longer lead times like 30 days, can be attributed to its hybrid approach, which combines physical principles with data-driven methods. This integration allows the model to better capture the underlying physical processes influencing SST, leading to more robust and reliable predictions as the forecast horizon extends. The ability to leverage both data patterns and physical laws makes the PINN-Transformer model more resilient to overfitting and better suited for extrapolating beyond short-term data trends, enhancing its long-term forecasting capabilities compared to purely data-driven models like CNN and LSTM. Overall, the PINN-Transformer model's strong performance highlights the advantages of incorporating physical knowledge into machine learning frameworks for complex environmental predictions.

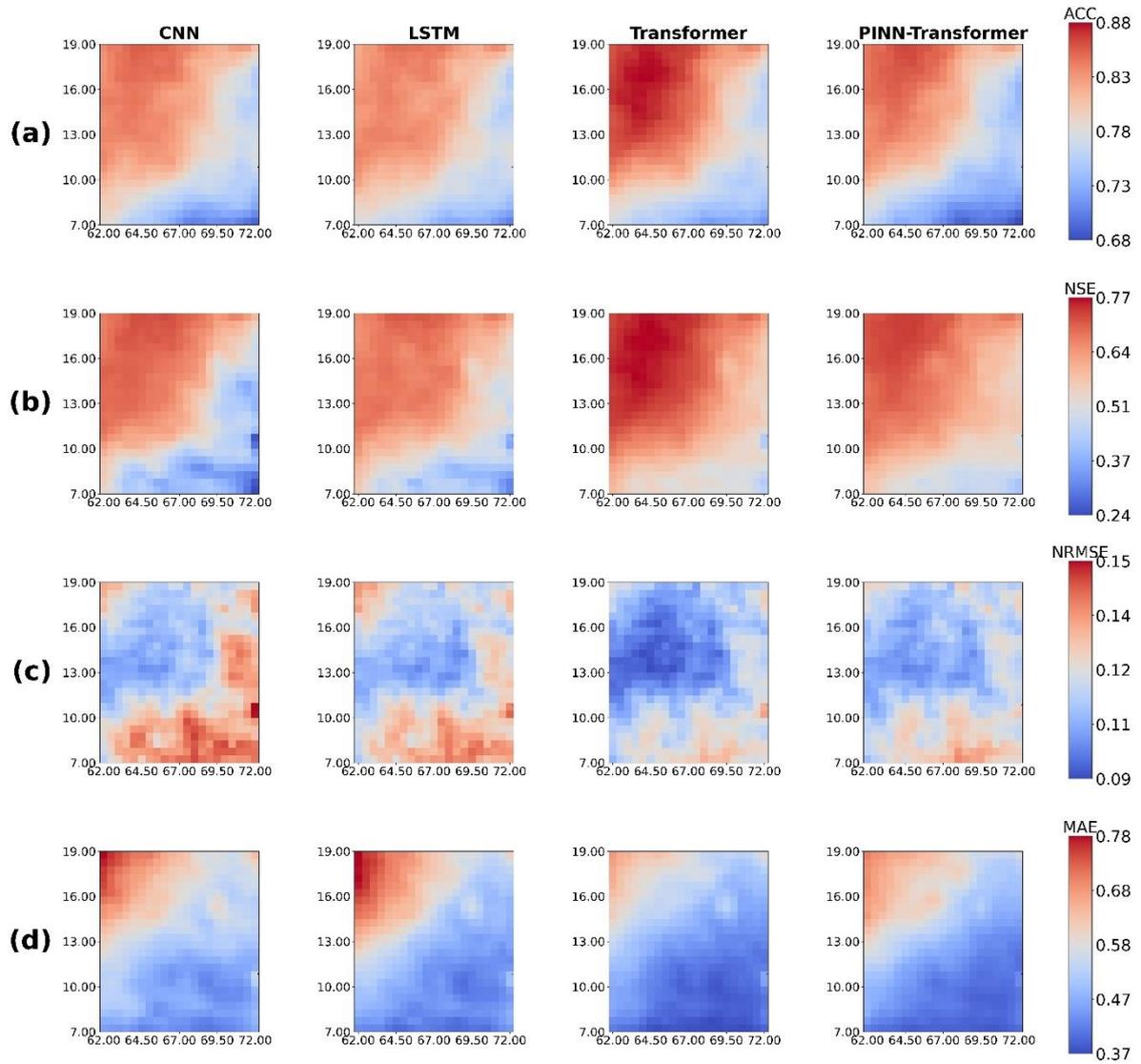

**Figure 8**. **(a)** CC; **(b)** NSE; **(c)** NRMSE; and **(d)** MAE values obtained during the testing period for CNN, LSTM, Transformer, and PINN-Transformer models for 30-day lead time.

Figure 9 presents scatter plots comparing observed and predicted SST across four models. The CNN model (Figure 9(a)) shows moderate NSE values, ranging from 0.34 to 0.53, indicating a fair correlation between observed and predicted SSTs, though with some variability. The LSTM model (Figure 9(b)) exhibits a similar range of NSE values, from 0.39 to 0.55, reflecting comparable accuracy to the CNN model. The Transformer model (Figure 9(c)) performs slightly better, with NSE values between 0.50 and 0.59, indicating improved predictive accuracy. The PINN-Transformer model (Figure 9(d)) demonstrates the highest NSE values, ranging from 0.46 to 0.58, indicating the strongest correlation and predictive efficiency among the models. This suggests that the PINN-Transformer model is particularly effective in maintaining alignment between observed and predicted values over a longer forecast period, likely due to its integration of physical principles with machine learning techniques. Overall, Figure 9 underscores the superior performance of the PINN-Transformer model in forecasting SST, particularly at extended lead times.

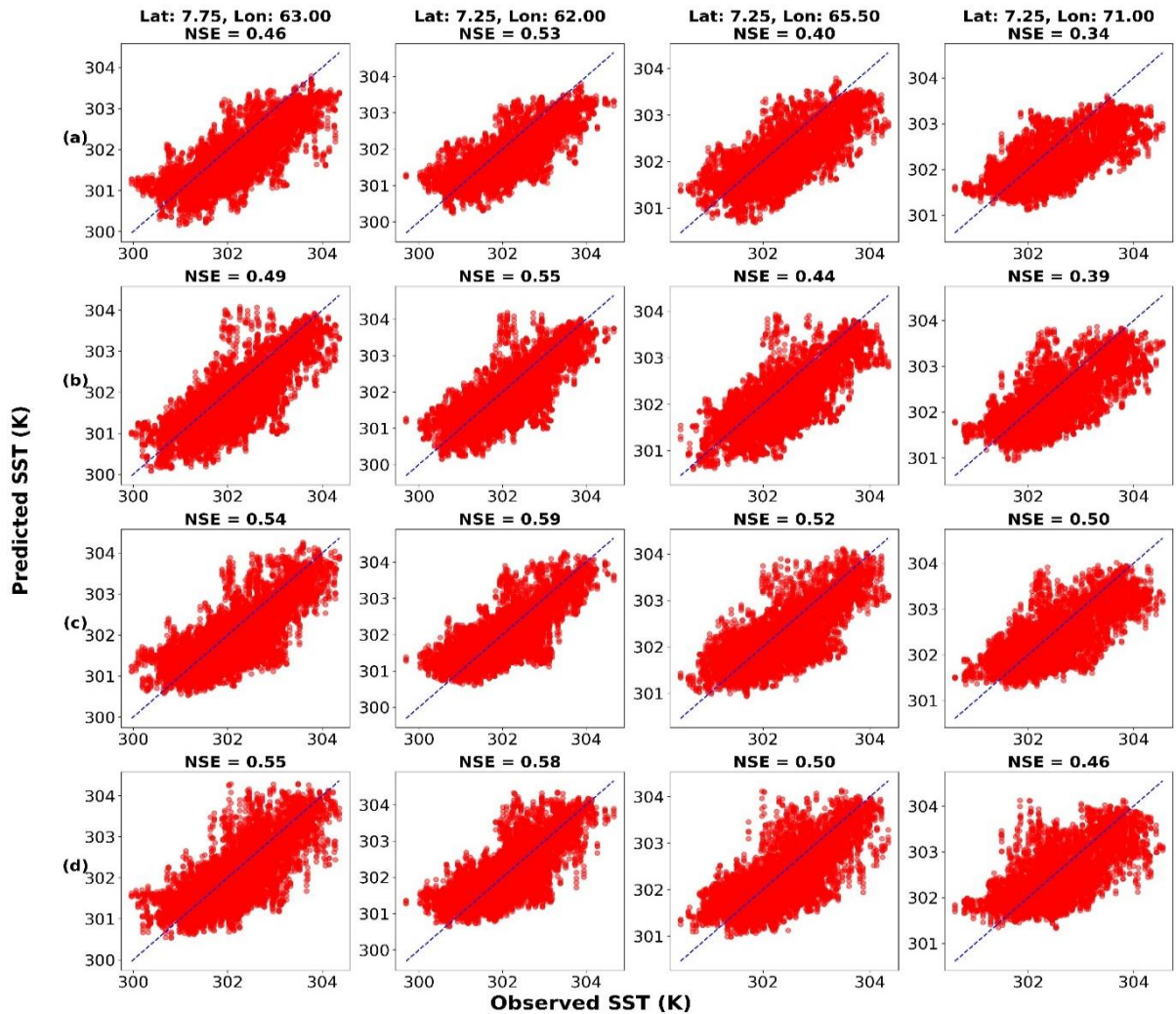

**Figure 9**. Scatter plot between observed and predicted SST for 30-day lead time using different models: **(a)** CNN; **(b)** LSTM; **(c)** Transformer; and **(d)** PINN-Transformer.

### 4.4. Discussion

Table 2 shows the range of ACC, NSE, NRMSE, and MAE obtained during the testing period for different lead times across the four models considered in this study. The percentage of grids with ACC > 0.8, NSE > 0.7, NRMSE < 0.1, and MAE < 0.5 are also presented in Table 2. For the 7-day lead time, the CNN, LSTM, and Transformer models perform very well. All three models achieve 100% of grids with ACC > 0.8. More than 95% of grids have NSE > 0.7,

NRMSE < 0.1, MAE < 0.5 in all these three models indicating highly accurate predictions. The PINN-Transformer model performs slightly less accurately, with 95.24% of grids having ACC > 0.8, 55.36% having NSE > 0.7, 46.23% having NRMSE < 0.1, and 79.37% having MAE < 0.5. This suggests that for short-term SST predictions, the CNN, LSTM, and Transformer models effectively capture the relevant spatial and temporal patterns (Zhang et al., 2017). However, the slightly lower performance of the PINN-Transformer at shorter lead times may be due to the introduction of unnecessary complexity from integrating physical principles, where purely data-driven approaches are sufficient for accurate predictions (Willard et al., 2022).

At the 15-day lead time, performance drops slightly for each model. The CNN model remains the best among all the models, with 99.60% of grids having ACC > 0.8 and 90.87% having MAE < 0.5. However, the percentage of grids with NSE > 0.7 and NRMSE < 0.1 drops below 85%, whereas it was above 95% for the 7-day lead time. The LSTM model shows 92.26% of grids with ACC > 0.8 and 84.13% with MAE < 0.5. Despite having over 80% of grids meeting ACC and MAE thresholds, the percentage drops to around 64% for NSE > 0.7 and 55% for NRMSE < 0.1. At the 7-day lead time, the LSTM model performed better compared to the Transformer model. However, with an increase to a 15-day lead time, the Transformer model outperforms the LSTM in terms of the percentage of grids with NSE > 0.7 and MAE < 0.5, with values around 65% and 61%, respectively. The PINN-Transformer model continues to perform the worst among the four models at the 15-day lead time, but the percentage of grids with NSE > 0.7, NRMSE < 0.1, and MAE < 0.5 increased to around 61%, 53%, and 83%, respectively, compared to the 7-day lead time. The PINN-Transformer's relative improvement at this lead time suggests the increasing importance of physical constraints as the lead time increases (Raissi et al., 2019).

For the 30-day lead time, the performance of all models declines considerably compared to the 7-day and 15-day lead times. For the CNN model, the percentage of grids with NSE > 0.7 and NRMSE < 0.1 was above 80% at both 7-day and 15-day lead times. However, at the 30-day lead time, these percentages drop to 12% for NSE and none of the grids meet the NRMSE < 0.1 threshold. The performance of the LSTM model is also poor compared to CNN, with only around 2% of grids having NSE > 0.7. The Transformer model performs better than the CNN, LSTM, and PINN-Transformer models, with around 20% of grids achieving NSE > 0.7, although this percentage was about 65% at the 15-day lead time. The performance of the PINN-

Transformer is also better compared to CNN and LSTM at the longer lead time, with a higher percentage of grids achieving NSE > 0.7.

**Tabel 2.** The range of Anomaly Correlation coefficient, Nash-Sutcliffe efficiency, Normalized root mean square error, and Mean Absolute Error values obtained during testing period for different lead time across four different models. The percentages of grids with ACC values >0.8, NSE values >0.7, NRMSE values < 0.1, and MAE value < 0.5 are also displayed.

| Lead Time | Models | Range | | | | Percentage of Grids | | | |
|---|---|---|---|---|---|---|---|---|---|
| | | ACC | NSE | NRMSE | MAE | ACC > 0.8 | NSE > 0.7 | NRMSE < 0.1 | MAE < 0.5 |
| 7-day | CNN | 0.86 - 0.96 | 0.68 - 0.92 | 0.06 -0.11 | 0.24 - 0.44 | 100.00 | 99.60 | 99.01 | 100.00 |
| | LSTM | 0.87 - 0.96 | 0.75 -0.92 | 0.06 -0.11 | 0.25 - 0.46 | 100.00 | 100.00 | 98.02 | 100.00 |
| | Transformer | 0.85 - 0.96 | 0.64 - 0.92 | 0.06 -0.10 | 0.28 - 0.42 | 100.00 | 95.63 | 97.82 | 100.00 |
| | PINN-Transformer | 0.77 - 0.91 | 0.38 - 0.81 | 0.08 -0.13 | 0.36 - 0.64 | 95.24 | 55.36 | 46.23 | 79.37 |
| 15-day | CNN | 0.80 - 0.92 | 0.57 - 0.84 | 0.07 -0.12 | 0.30 - 0.59 | 99.60 | 82.34 | 83.73 | 90.87 |
| | LSTM | 0.74 - 0.92 | 0.50 - 0.81 | 0.08 - 0.13 | 0.32 - 0.61 | 92.26 | 64.09 | 54.96 | 84.13 |
| | Transformer | 0.77 - 0.93 | 0.53 - 0.86 | 0.07 - 0.13 | 0.34 - 0.52 | 95.24 | 65.28 | 61.11 | 98.02 |
| | PINN-Transformer | 0.75 - 0.91 | 0.48 - 0.82 | 0.08 - 0.13 | 0.34 -0.59 | 87.50 | 61.51 | 53.37 | 83.33 |
| 30-day | CNN | 0.70 - 0.86 | 0.24 - 0.73 | 0.10 - 0.15 | 0.42 - 0.78 | 55.36 | 11.90 | 0.00 | 36.51 |
| | LSTM | 0.71 - 0.85 | 0.31 - 0.71 | 0.09 - 0.15 | 0.40 - 0.78 | 58.13 | 2.38 | 0.20 | 51.79 |
| | Transformer | 0.72 - 0.88 | 0.43 - 0.77 | 0.09 - 0.14 | 0.37 - 0.68 | 58.13 | 35.91 | 20.04 | 65.48 |
| | PINN-Transformer | 0.68 - 0.87 | 0.41 - 0.74 | 0.08 - 0.14 | 0.38 - 0.70 | 47.02 | 19.25 | 0.20 | 55.16 |

To assess the model's performance, the maximum observed SST value for each grid is first determined. Subsequently, the difference between the observed SST and the predicted SST values is calculated for the day corresponding to this maximum observed SST value. This approach ensures that the comparison is made specifically on the day when the observed SST reaches its peak. This difference is plotted in Figure 10 for all the four models at different lead time. In Figure 10(a) for the 7-day lead time, the CNN, LSTM, and Transformer models exhibit predominantly positive anomalies (in red) across most grids, indicating accurate SST predictions. The PINN-Transformer model, however, shows a mix of positive and negative anomalies (in blue), suggesting less consistent performance in capturing SST patterns compared to the other models. In Figure 10(b) for the 15-day lead time, all models continue to show mostly positive anomalies, but there is a noticeable increase in the presence of negative anomalies. This indicates a slight decrease in prediction accuracy as the lead time increases. The PINN-Transformer model again shows a more varied pattern, with more pronounced negative anomalies compared to the other models.

In Figure 10(c) for the 30-day lead time reveals the most significant differences in model performance. The CNN, LSTM, and Transformer models show a marked increase in negative anomalies, indicating a further decline in prediction accuracy over longer lead times. The PINN-Transformer model continues to display a mix of anomalies, with prominent negative anomalies, suggesting that the physical constraints integrated into this model may be affecting its performance over longer lead times. Overall, the Figure 10 illustrates that while the CNN, LSTM, and Transformer models maintain relatively consistent anomaly patterns for short lead times, their accuracy decreases as the lead time extends to 30 days. The PINN-Transformer model shows more variability in its anomaly patterns across all lead times, reflecting the influence of physical constraints and their potential impact on prediction accuracy.

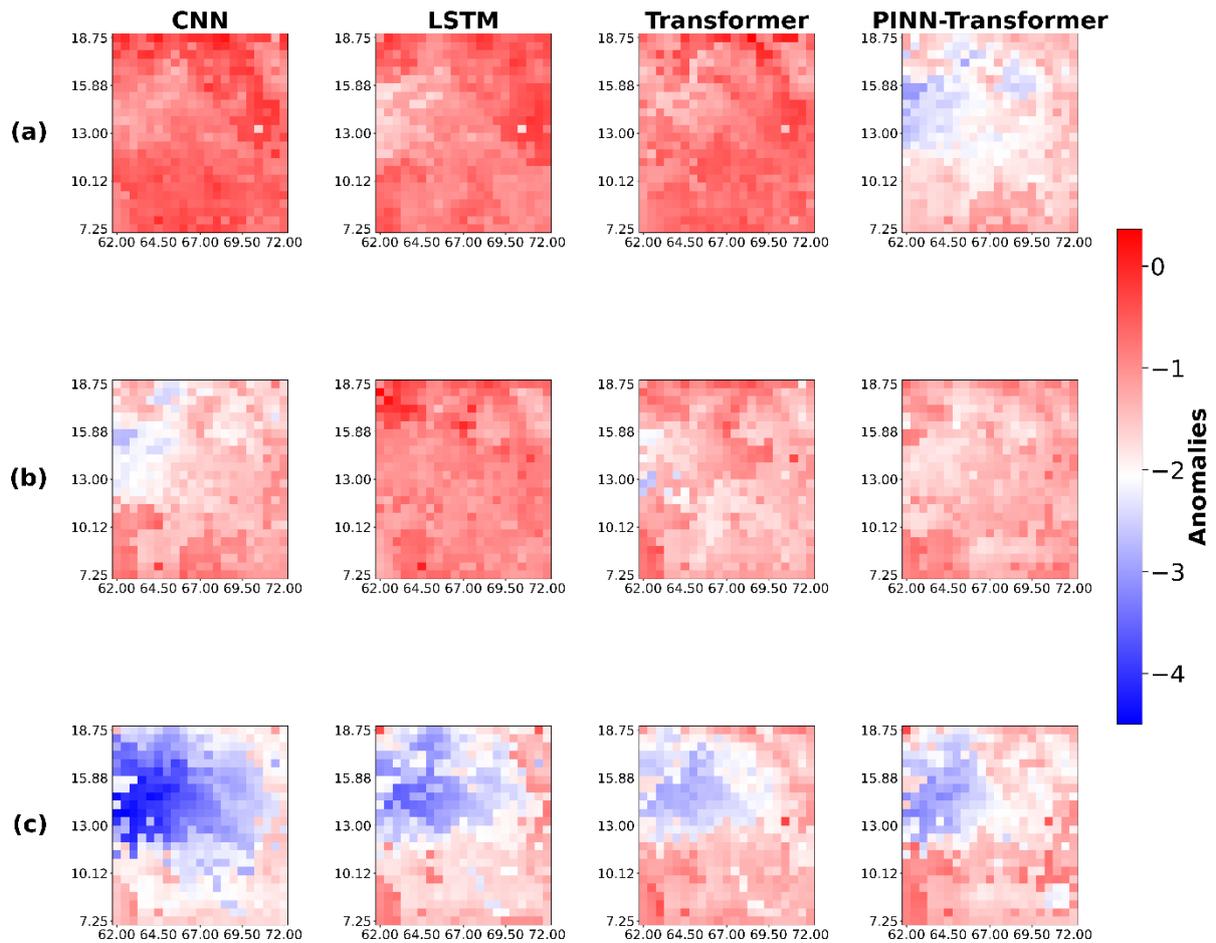

**Figure 10.** Maximum SST anomalies: **(a)** 7-day; **(b)** 15-day; and **(c)** 30-day lead time predictions obtained during the testing period for CNN, LSTM, Transformer, and PINN-Transformer models.

Similar to the maximum value, the minimum SST anomaly is also determined at each grid and is shown in Figure 11 for all models considered in this study across different lead times. In Figure 11(a) for the 7-day lead time, the CNN, LSTM, Transformer, and PINN-Transformer models exhibit predominantly negative anomalies (in blue) across most grids, indicating underpredictions of SST. The PINN-Transformer model shows a mix of positive anomalies (in red), suggesting some regions where it overpredicts SST. In Figure 11(b) for the 15-day lead time, all models start to show more pronounced positive anomalies, indicating areas where the models overpredict SST. The LSTM and PINN-Transformer models exhibit larger regions of positive anomalies compared to the CNN and Transformer models, indicating higher overprediction tendencies at this lead time.

Figure 11(c) for the 30-day lead time shows a further shift in anomaly patterns. The CNN and Transformer models exhibit a mix of positive and negative anomalies, indicating a balanced yet less accurate prediction. The LSTM and PINN-Transformer models continue to show larger regions of positive anomalies, reflecting increased overprediction at this longer lead time. Overall, Figure 11 illustrates that while the CNN and Transformer models maintain relatively consistent anomaly patterns for shorter lead times, their accuracy decreases and shows a mix of underprediction and overprediction at the 30-day lead time. The LSTM and PINN-Transformer models exhibit more variability, with significant overprediction as the lead time extends, reflecting the challenges these models face in maintaining accuracy over longer prediction horizons.

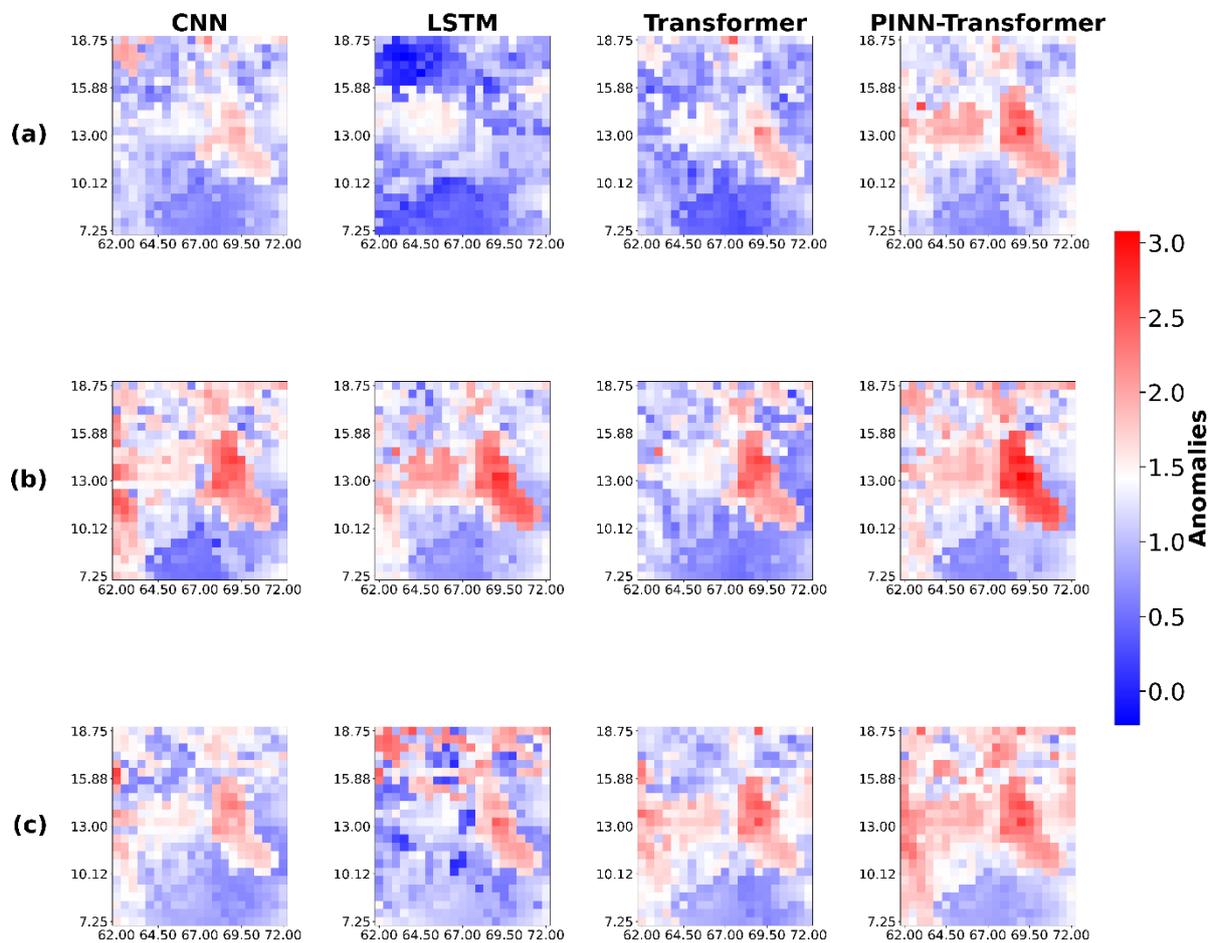

**Figure 11**. Minimum SST anomalies: **(a)** 7-day; **(b)** 15-day; and **(c)** 30-day lead time predictions obtained during the testing period for CNN, LSTM, Transformer, and PINN-Transformer models.

These findings have important implications for the field of SST prediction and, more broadly, for oceanic and climate modeling. The varying performance of models across lead times underscores the need for a nuanced approach to model selection based on the specific requirements of the forecasting task. For operational short-term SST forecasting, CNNs or LSTMs may be preferred due to their high accuracy and computational efficiency. However, for medium to long-term predictions, especially in applications where error minimization is critical, Transformer-based models, particularly those incorporating physical constraints, emerge as more suitable choices. In the broader context of climate science and oceanography, this study underscores the potential of deep learning approaches in enhancing our understanding and prediction of complex oceanic phenomena. The ability to accurately forecast SST patterns, particularly over extended periods, has significant implications for climate modeling, marine ecosystem management, and weather prediction.

## 5. Conclusions

Recently, machine learning has demonstrated its utility in predicting various atmospheric and oceanic variables. In this study, we test the efficacy of different state-of-the-art machine learning algorithms in predicting SST and examine how physics-informed machine learning enhances prediction accuracy. Specifically, we evaluate these algorithms for short-term forecasting across 7-day, 15-day, and 30-day lead times. We test the performance of four models—CNN, LSTM, Transformer, and PINN Transformer—using ERA5 data at a spatial resolution of 0.5° x 0.5°. The performance of these models was assessed during both the training and testing phases using ACC, NSE, NRMSE, and MAE as the statistical metrics. In the 7-day forecast, all models performed well, while for the 15-day lead forecast, the Transformer model exhibited the best performance. Both the CNN and Transformer models exhibited robust performance in the 30-day forecast. Significantly, when considering longer forecast periods, the combination of physical principles with machine learning, as exemplified by the PINN Transformer, yielded a distinct advantage, resulting in superior performance compared to alternative models. Ultimately, when it comes to short-term sea surface temperature (SST) projections spanning from 7 to 15 days, models that depend solely on ML without incorporating physical integration generally exhibit superior performance. However, when it comes to making predictions for a longer period of time, models that include physical principles, such as the PINN Transformer, demonstrate superior performance compared to other models. This suggests that although the incorporation of physical integration may not enhance short-term forecasts, it is essential for enhancing the precision of longer-term projections.

While the findings of this study are promising, there are certain limitations to note. First, only four models—CNN, LSTM, Transformer, and PINN-Transformer—were evaluated. Including a broader range of models in future comparisons could help identify the most effective approach. Second, the study focused on six atmospheric variables, but incorporating additional factors, such as wind shear or ocean currents, may improve the models' learning potential. Third, due to computational limitations, the analysis was restricted to 504 grids at a 1-degree spatial resolution. Expanding the spatial domain and enhancing resolution in future studies could lead to better modelling of finer-scale SST features and interactions. Future research should prioritize testing a wider variety of models, adding more atmospheric variables, and enhancing computational efficiency to support higher-resolution data and larger study areas.

**Data availability**

The original observational data are publicly available. All data sources are mentioned in the Materials and Methods section. The source codes for the analysis of this study are available from the corresponding author upon reasonable request.

**Declarations**

**Competing interests:** The authors declare that they have no competing interests.